# Network Slicing-based Customization of 5G Mobile Services

Ibrahim Afolabi, Tarik Taleb, Pantelis A. Frangoudis, Miloud Bagaa and Adlen Ksentini

*Abstract*—Through network slicing, different requirements of different applications and services can be met. These requirements can be in terms of latency, bandwidth, mobility support, defining service area, as well as security. Through fine and dynamic tuning of network slices, services can have their delivery platforms constantly customized according to their changing needs. In this article, we present our implementation of an E2E network slice orchestration platform, evaluate its performance in terms of dynamic deployment of network slices in an E2E fashion, and discuss how its functionality can be enhanced to better customize the network slices according to the needs of their respective services.

## I. Introduction

The fifth generation of mobile communications system (5G) is soon to be deployed, supporting the business requirements of different mobile virtual network operators (MVNO), over-thetop (OTT) application providers, and vertical industries [1]. 5G is expected to positively impact and revolutionize the Quality of Service (QoS) perceived by users through an agile and truly programmable network architecture that allows a genuine service differentiation. A wide gamut of applications is expected to be provided by 5G. These applications have different requirements and pose different levels of challenges to the underpinning mobile network infrastructure [2]. Their requirements stem from the individual characteristics exhibited by each of the mobile network services belonging to the MVNO, OTT or vertical industry, operating on top of the same physical network infrastructure.

Most importantly, the 5G vertical industries (e.g., health care, automotive, media and entertainment, and smart manufacturing) have demonstrated the need for distinctive service differentiation. Such service differentiation can be enabled by accommodating the preferences of end-users through the customization of the underlying service delivery networks [3]. For instance, in case of the media and entertainment vertical, service customization should go beyond servicing content selected only as per the preferences (e.g., movie types and preferred quality) and contextual information (e.g., location, language and ethnicity) of viewers, but should also reflect the network dynamics and the mobility features of viewers. For example, a video content provider could recommend a video content to a group of viewers based on their spoken language. Knowing in addition the mobility patterns of these viewers, the video content provider could in advance arrange for the right amount/slice of network resources over the areas to be visited by the viewers so that the viewers' perceived Quality of Experience (QoE) would be maintained at a certain desired level. In this fashion, service customization goes beyond users' preferences and contextual information, to also build optimal network slices to efficiently deliver the services as well. Accordingly, service customization becomes a multidimensional concept, whereby both the users' context and the underpinning network's context (e.g., available network resources, application resources, available technology, and available network function types) are all taken into consideration when orchestrating a particular network slice to serve best the overall needs of a 5G vertical.

Ibrahim Afolabi, Tarik Taleb, and Miloud Bagaa are with the Department of Communications and Networking, School of Electrical Engineering, Aalto University, Finland (e-mails: {firstname.lastname}@aalto.fi). Tarik Taleb is also with the Faculty of Information Technology and Electrical Engineering, Oulu University, and with the Department of Computer and Information Security, Sejong University, Seoul 05006, South Korea.
Pantelis A. Frangoudis and Adlen Ksentini are with the Communication Systems Department, EURECOM, Sophia Antipolis, France (e-mails: {firstname.lastname}@eurecom.fr).

By enabling promising 5G-ready applications as well as complex systems with diverse service and network requirements on top of a shared infrastructure through network slicing, 5G technology will address the needs of all network users in parallel.

In this article, we present a novel framework for the dynamic orchestration of E2E network slices. We also present an evaluation of the system, and identify a number of useful information for 5G network and infrastructure providers. Our slice orchestration framework supports the creation of network slices on top of shared infrastructure, and customized to serve different 5G mobile services. The proposed framework is compliant with 3GPP specifications on network slicing orchestration and management [4].

The remaining of this article is organized in the following fashion. Section II gives a comprehensive overview of network slicing, describing slice components, types and templates. Section III presents our architectural framework for E2E network slicing, focusing on its distinctive characteristics. Section IV evaluates the performance of our envisioned framework from user- and system-centric perspectives. The article concludes in Section V.

## II. Network Slicing

Network slicing is the process of sharing network resources, such as computing, networking, memory and storage which are TABLE I: Table of Abbreviations

| Acronyms | Full meanings |
|---|---|
| 3GPP | 3rd Generation Partnership Project |
| 5G | 5th Generation |
| CN | Core Network |
| CSMF | Communication Service Management Function |
| E2E | End-to-End |
| MAC | Medium Access Control |
| mMTC | Massive Machine Type Communication |
| MVNO | Mobile Virtual Network Operator |
| NBI | Northbound Interface |
| NFV | Network Function Virtualization |
| NFVI | NFV Infrastructure |
| NFVO | NFV Orchestrator |
| NSMF | Network Slice Management Function |
| NSSMF | Network Slice Subnet Management Function |
| MANO | Management and Orchestration |
| MME | Mobility Management Entity |
| OTT | Over-The-Top |
| PHY | physical |
| QoE | Quality of Experience |
| QoS | Quality of Service |
| RAN | Radio Access Network |
| RRB | Radio Resource Blocks |
| SDN | Software Defined Networking |
| uRLLC | Ultra Reliable Low Latency Communication |
| vEPC | Virtual Evolved Packet Core |
| VIM | Virtualized Infrastructure Manager |
| VM | Virtual Machine |
| VNF | Virtual Network Function |
| xMBB | Extreme Mobile Broadband |

available on a single set of physical infrastructures, among different virtual tenants with a certain degree of logical or physical separation as detailed in [5].

### *A. Slice Components, Types and Templates*

*a) Slice components:* In general, a network slice consists of a number of VNFs interconnected and optimally placed to fulfill a specific set of requirements and meet specific network constraints in order to realize a particular use-case scenario. Since every network slice is designed to deliver a specific usecase scenario, the set of VNFs needed to create an instance of a network slice will be somewhat different from another. However, despite the possible differences in the design of network slices, 5G technology aims to incorporate simplicity and flexibility in network slicing. This will be done by sharing and reusing as many network functions as possible in the instantiation of different network slices [6].

*b) Slice types:* In the context of E2E network slicing in 5G, a network slice will consist of components in the form of VNFs, from the Radio Access Network (RAN), the Core Network (CN), and the transport network.

According to [7], network slices are broadly classified into three major categories: Extreme/Enhanced Mobile BroadBand (xMBB), Ultra Reliable Low Latency Communication (uRLLC) and Massive Machine Type Communications (mMTC). Each of these categories is characterized by its specific performance requirements, which are reflected through the VNF types and the amount of virtual resources (i.e., CPU, Storage, Memory) used.

*c) Slice templates:* The aforementioned categorization of network slices facilitates the definition, design and optimization of blueprints for different network slices belonging to the different categories. These variant network slice blueprints are known as *network slice templates*. A network slice template is a functional blueprint from which a particular type of network slice could be instantiated. The blueprint defines all the necessary VNFs, their composition, VNF forwarding graph, virtualization technology type, location(s) of instantiation, level of elasticity and E2E network resources needed by a network slice. In addition, the slice's lifespan (i.e., the operational duration of the slice) is determined through its blueprint by the values given for its activation and termination times (e.g., in day(s), week(s), month(s) or even in year(s)), respectively.

### *B. End-to-End Network Slice Structure*

At a high level, regardless of the underlying system and the functional requirements of a network slice and its VNF composition, an E2E network slice should always be composed of three sub-slices: RAN, Core, and Transport.

*1) RAN sub-slice:* Ensuring that a network slice is E2E implies that the RAN must be also sliceable. A notable challenge here is how to provide the necessary level of performance isolation across slices which are sharing typically the same access network, where each slice has its own resource requirements. These requirements are basically defined by the available amount of the physical radio resource blocks (RRBs) and the frequency of scheduling them (determining the slice's bandwidth and latency). This physical resource could either be statically or dynamically allocated [8] at the access network for use by the RAN sub-slices.

The isolation between RAN sub-slices should be ideally carried out across the protocol layers responsible for the aforementioned scheduling and physical radio resource allocation; particularly, the PHY and MAC scheduling layers of the mobile access node [9]. The RAN sub-slicing could be enabled for instance by assigning each sub-slice a unique ID [6]. This ID is used to identify the RAN sub-slices and is also used to enforce RAN-level differentiated traffic treatment per subslice.

*2) Core sub-slice:* This sub-slice includes the elements that correspond to the core network. These elements naturally lend themselves to a virtualized implementation.

NFV facilitates tailoring a slice to the needs of its respective service. Allowing the deployment of a virtualized core network over an NFV Infrastructure (NFVI) comes with the flexibility to customize the compute and other resources allocated to the slice. It also allows the slice to scale on demand for cost and performance optimization, and to select the appropriate functional configuration of the core network components.
These tasks are supported by maturing relevant standards, such as the ETSI NFV Management and Orchestration (NFVMANO) [10] framework and available implementations of MANO software stacks.

*3) Transport sub-slice:* After instantiating a virtual core network sub-slice from a slice template, the VNFs constituting the core sub-slice should be chained together. The resulting core sub-slice will then be connected to the appropriate RAN sub-slice and then to an external network through a corresponding transport path. The involved transport path will form the transport sub-slice, whose management is enabled by the use of technologies such as SDN or Virtual LANs.

All sub-slices provide a communication service (service slice) at different technological domains. In addition, they include the necessary dedicated resources (resource slice) needed to operate the technology-specific service. For instance, at the RAN level, the service is radio access and the resources are virtual resource blocks. The core sub-slice provides EPCaaS, which is supported by specific dedicated virtual storage and compute resources. The transport sub-slice provides a connectivity service to external networks, with dedicated virtual network links.

## III. An End-to-End Slice Orchestration and Management Framework

### *A. High-level design*

In compliance with the 3GPP specifications of [4], our high-level network slicing framework is shown in Fig. 1. The Communication Service Management Function (CSMF) component, which is outside the scope of this work, receives a request for a communication service by a vertical and translates it into specific slice requirements. Then, it interacts with the *End-to-End (E2E) Slicer* via the latter's northbound interface (NBI) to request the instantiation of an E2E network slice with specific characteristics. The features, design, and implementation of the E2E Slicer are the main focus of this article. For more details on CSMF functions, the interested reader may refer to [4].

Every 5G service slice shall have a lifecycle, which is handled by the E2E Slicer's Network Slice Management Function (NSMF). In particular, based on the requested slice type and specific dimensioning information as reflected in the slice's requirements (e.g., slice reliability targets, number of UEs, traffic characteristics to handle, etc.), a suitable slice template shall be retrieved from a catalogue and appropriate Network Slice Subnet Management Functions (NSSMF) that correspond to each sub-slice shall be selected, as indicated in the template. The NSMF then proceeds by composing a customized slice instance and delegating the instantiation and management of each sub-slice to the appropriate NSSMF. Note that our framework supports the *MANO-as-a-Service* concept by design. Effectively, each slice template includes fields that indicate the type of NSSMF that should be used to orchestrate the underlying sub-slice instance. This functionality is particularly important for handling core network sub-slices. In this case, the NSMF will identify which NFV Orchestrator (NFVO) and the respective MANO stack to launch or reuse one slice instantiation type. Then, it will access the NFVO's NBI to request the creation of the appropriate core network sub-slice instance, appropriately customizing the included VNF instances (e.g., VMs implementing the Mobility Management Entity (MME) functionality) and the respective resources according to the specific service characteristics.

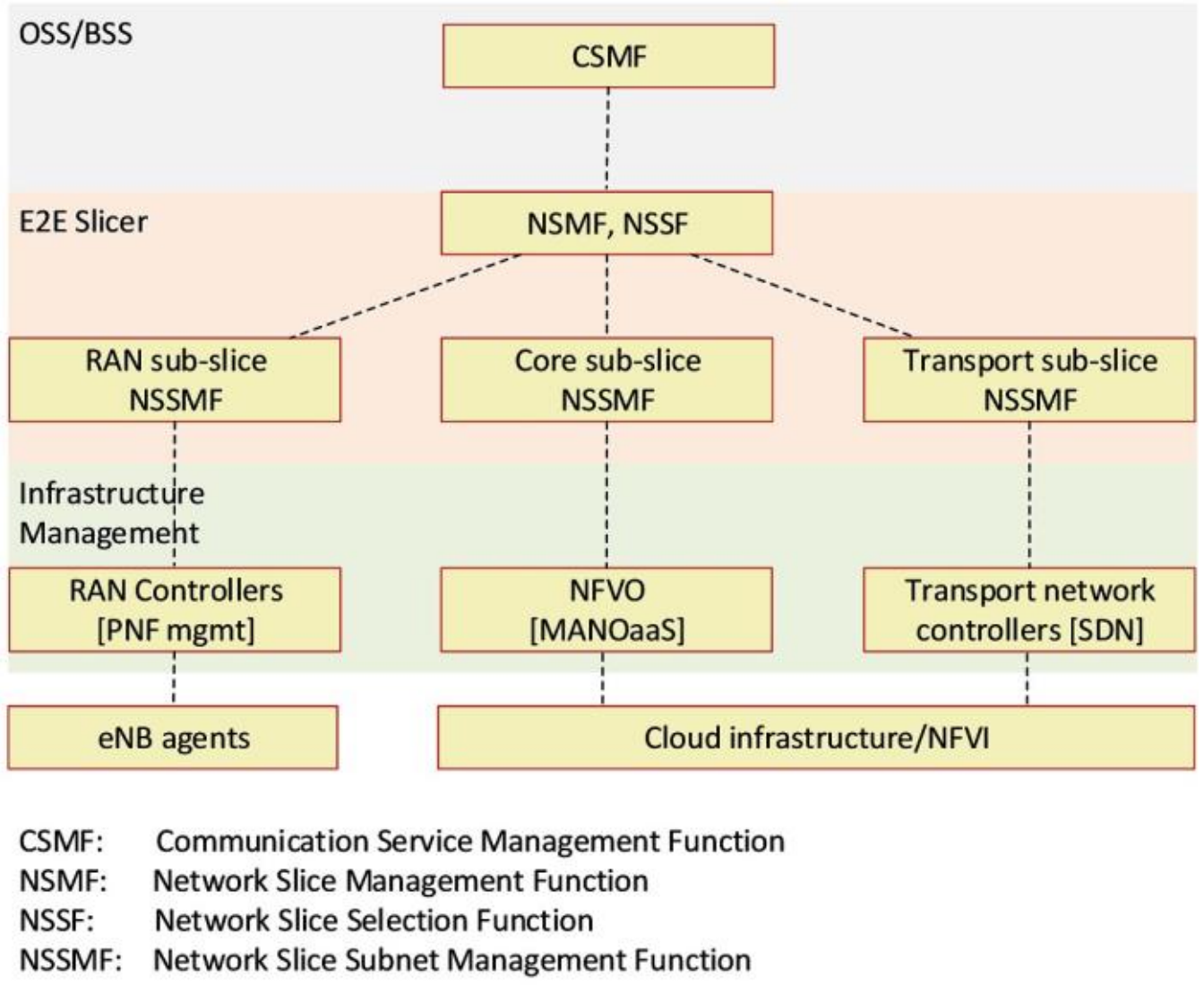

Fig. 1: High-level view of our E2E network slicing architecture. Our design features an E2E Slicer which implements NSMF and NSSF components. It supports the integration of PNFs for RAN slicing and different MANO stacks per slice type and interacts with them via different NSSMF components per sub-slice.

The RAN sub-slice NSSMF includes a RAN resource allocator component, whose role is to translate slice requirements into a radio resource allocation and carry out high-level RAN resource management. As a result, the RAN's state needs to be always maintained in terms of the connected UEs per eNodeB and the quality of their radio connection, per UE/slice bitrate requirements, and the slice instances to which an eNodeB participates. The above information is used by the NSSMF to derive an appropriate RAN resource partition per cell in order to satisfy the requirements of the coexisting slices (e.g., latency). The RAN resource partitioning can be adjusted dynamically, following updates in the RAN state. Changes in the radio conditions or the deployment of new slices is enforced by the NSSMF using the appropriate NBI of the RAN controller. In this context, the RAN controller can be seen as a RAN-specific Virtual Infrastructure Manager (VIM), while an agent installed at an eNodeB works as a hypervisor: It exposes a virtualized view of RAN resources and offers the necessary primitives to execute resource management tasks across slicededicated physical resources on the same radio hardware.

Finally, the role of the transport-level NSSMF is to interact with network elements such as SDN controllers, in order to manage the provisioning and isolation of the links connecting (virtual or physical) network functions of the access and core network, and towards external networks.

A design with a similar target but distinct differences to our approach is JOX [11], a Juju-based slicing-oriented orchestration scheme. Although the design of JOX does not preclude an NSMF operation, it is more oriented towards providing NSSMF functionality and does not address slice selection issues. As such, it can be used in a complementary fashion to our architecture. In such a case, JOX could be used to implement different NSSMF plugins, e.g., abstracting the operation of the RAN controller by offering a RANspecific NSSMF accessible via its NBI. Similar to this is the Open Source MANO (OSM), which is a general orchestration framework that offers lots of features but with no native support for orchestrating and managing PNFs from which the RAN sub-slices are orchestrated at the eNodeB. Another similar project is M-CORD [12] which has made significant contributions to E2E network slicing. Based on its architecture, M-CORD includes slice definition components and allows stitching RAN and core network sub-slices as in our case. Given these similarities at the architectural and functional level, it should be noted that our focus is more on ensuring the compliance of our design with the 3GPP specifications on network slicing, and on describing our own technical solutions for RAN slicing, slice-dedicated core network support, and network slice selection and orchestration, in a more lightweight design and implementation than M-CORD.

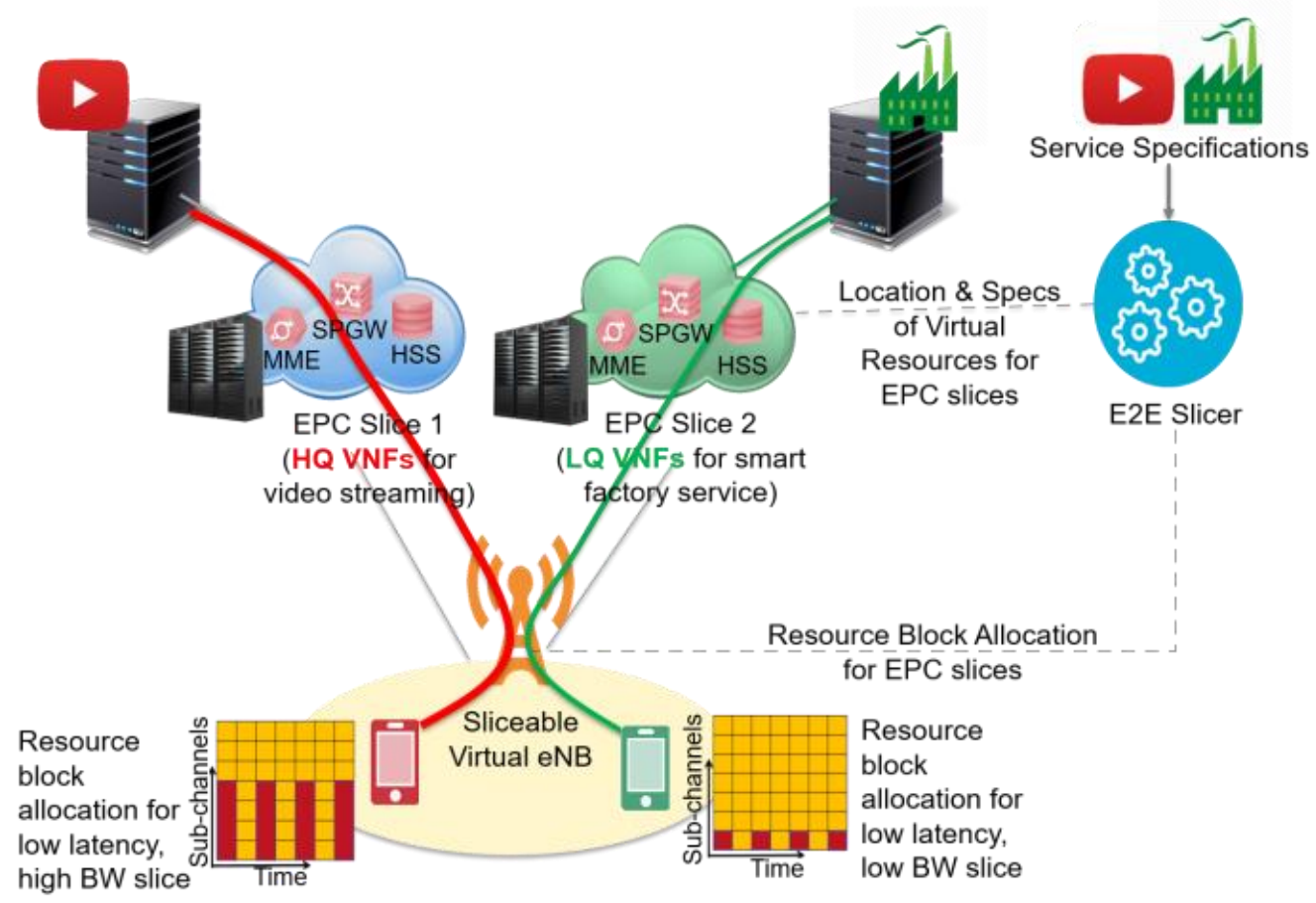

Fig. 2: 5G vertical use cases over the envisioned network slicing architecture.

### B. Use cases

Different vertical industries have different slicing requirements, as depicted in Fig. 2. In particular, we focus on the media and entertainment and the smart industry use cases and show how our E2E Slicer customizes and orchestrates two E2E network slices based on both system requirements as well as customer preferences. For example, in the case of the smart factory slice, shown in green, the service is deemed as a "low latency, low bandwidth" service. Effectively, information transmission of smart devices running in a smart factory would require very short latency as well as moderately small bandwidth at the access network. The access network resources are accordingly allocated in the form of one RRB (i.e., due to the "low bandwidth" nature of the smart factory service) to the slice and that is at certain predefined times corresponding to the "low latency" nature of the smart factory service. Correspondingly, the VNFs that would be instantiated at the core network would be more lightweight (i.e., also of low quality) since the traffic to be processed at the control and data planes is not resource demanding.

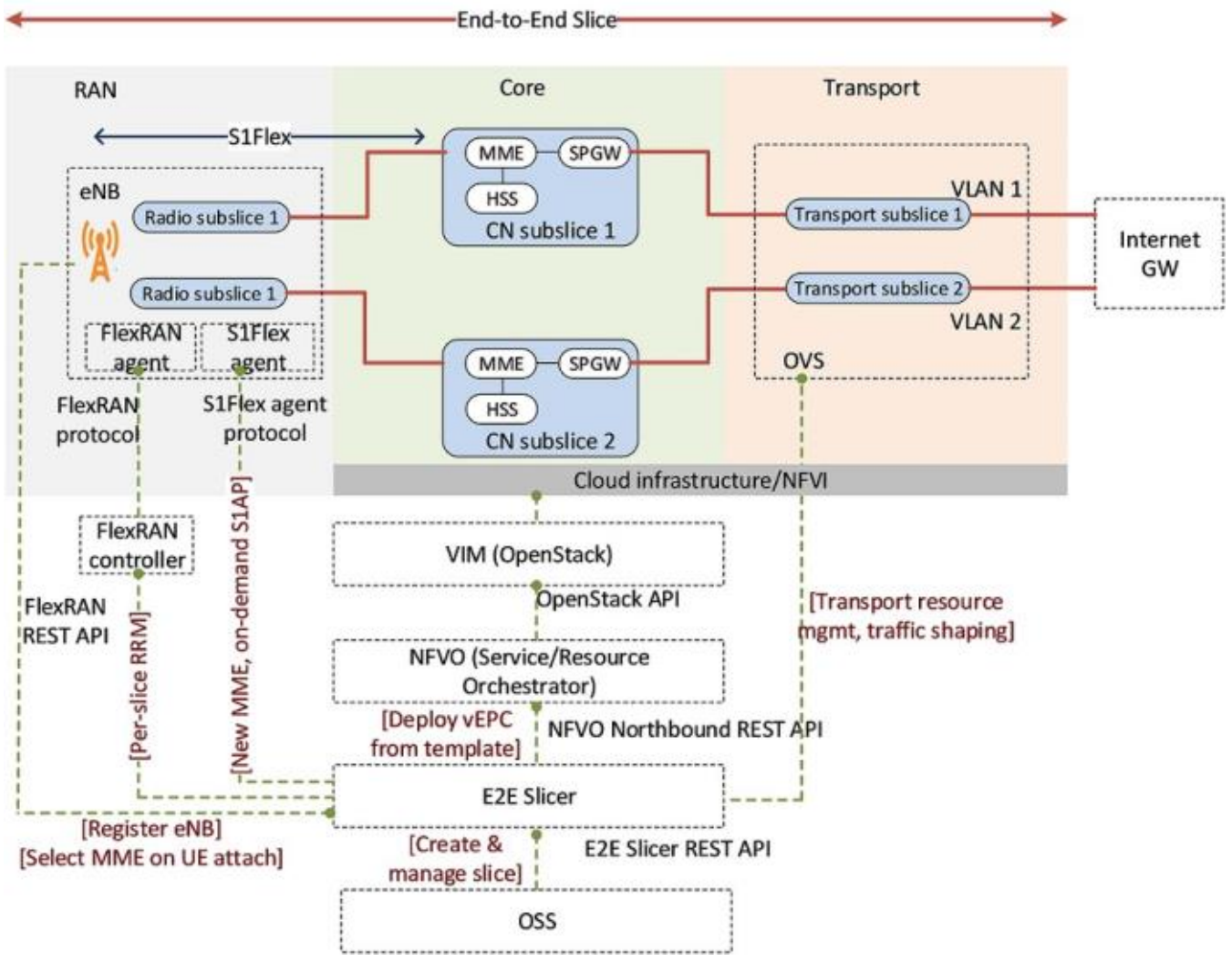

Fig. 3: Detailed architecture of the proposed framework.

Conversely, in the case of the slice dedicated to a video streaming service shown in red, the service is deemed as "low latency, high bandwidth". Consequently, the corresponding amount of network resources allocated at both the core and access parts of the network is higher. Many more resource blocks are frequently allocated at the RAN compared to the smart factory use case. Similarly, the VNFs instantiated for the core network sub-slice are of high quality, able to process video packets and are running on more virtual resources.

### C. Distinctive features

In this section, we present some of our distinctive design and implementation choices for the E2E Slicer. Our detailed architecture, following the high-level design principles of Section III-A, is shown in Fig. 3.

It shall be noted that our orchestration system supports instantiating and orchestrating network slices on multiple clouds/NFVOs. This functionality is provided by specific percloud NSSMFs. However, the scope of this work does not encompass the details of the individual functionalities and interfaces required to interact with each of the API endpoints provided by the different existing cloud administrative domains' NSSMFs. Moreover, the Orchestrated E2E slices are composed of sub-slices cutting across all the technology domains of the mobile network as presented in Fig. 3.

*1) RAN slicing:* To enable RAN slicing, we have adopted the FlexRAN approach [13]: Each eNodeB runs an agent which is managed by a centralized controller using a southbound protocol. The agent exposes RAN-level status information and, importantly, makes it possible to apply resource sharing policies at the MAC layer. It is left to the RAN subslice NSSMF to derive the appropriate resource allocation ratios across all coexisting slices sharing an eNodeB, and the FlexRAN protocol allows the NSSMF to enforce them using the appropriate API call.

Various algorithms [14] for deciding how RAN resources should be shared across slices are possible, and we do not discuss them here due to space limitations. Nevertheless, it is worth noting that these algorithms combine awareness of service-specific slice requirements (e.g., a strict latency constraint for a uRLLC slice or a high-throughput demand for xMBB) and the current RAN conditions. The former are provided to the E2E Slicer by the slice owner at slice instantiation time via its northbound REST interface, and the latter are retrieved periodically or on-demand by eNodeBs using FlexRAN.

*2) Slice-dedicated core network support:* A critical feature in order to support E2E slicing is the ability of an eNodeB to maintain associations with multiple core networks simultaneously, a feature typically termed as *S1-flex*. This is important since a shared eNodeB instance may be hosting multiple RAN sub-slices, which in turn are stitched with separate core sub-slices potentially running over different NFVIs. In our case, we have implemented a special S1-flex agent in OpenAirInterface[1], our LTE software of reference, that handles new eNodeB-MME associations (carried out using the S1AP protocol) when instructed so by the E2E Slicer. Notably, this feature not only enables the communication of an eNodeB with different slice-dedicated core networks, but also serves for load balancing purposes: A single slice may necessitate the instantiation of multiple replicas of its core network components (e.g., MME) to share the UE load; the E2E Slicer will have to use the same procedure to notify the involved eNodeBs about this.

*3) Network slice selection function:* Beyond the E2E Slicer's task of managing the association between RAN and core network sub-slices, it also has a central role in network slice selection. In traditional 4G networks, upon UE attachment, the eNodeB selects the MME responsible for handling that UE based on the specified PLMN (Public Land Mobile Network) identity included in the attach request. In our case, this decision is delegated to the slicer itself. In particular, each time a UE connects to an eNodeB, the latter accesses a special REST API endpoint of the E2E Slicer to query for the appropriate core sub-slice's MME to direct the UE to. This offers flexibility, since it facilitates the development of sophisticated per-slice core network selection policies implemented as E2E Slicer plugins, S1AP traffic load balancing algorithms, etc. However, as the required flexibility comes with centralizing slice selection decisions, it can lead to increased attachment latencies due to the additional eNodeB-slicer communication, but also potential overloads at the slicer level. In Section IV, we discuss this flexibility-performance trade-off.

*4) Slice instantiation and management:* The E2E Slicer provides a RESTful HTTP interface to the CSMF for slice instantiation and runtime management. This interface allows, among others, to select the type of slice from the ones available in the slice catalogue and continue with the composition of the slice as described in Section III-A.

Fig. 4 depicts the sequence of actions and the interactions between the components of our architecture that take place from the moment slice instantiation is requested until the slice is fully operational E2E. Note that we make the distinction between *default* and *dedicated* slices. In our design, the deployment of a default slice is necessary as a fall-back for the NSSF of the E2E Slicer when a UE is not associated with any specific slice. In a typical scenario, the core network sub-slice's NSSMF corresponding to the default slice may be simpler, in the sense that the default slice may involve an already running legacy core network, thus not requiring its dynamic deployment over a NFVI and its orchestration using a MANO stack. The discussion in this section focuses on the case for a *dedicated* slice, which in turn involves the dynamic instantiation of a slice-dedicated core network.

Apart from information on the type of the slice to deploy, a slice instantiation request also includes the involved service areas. It should be noted that the E2E slicer is aware of the eNodeBs belonging to each service area. This will enable it to configure the deployed RAN sub-slices appropriately. Based on the slice type as specified in the slice template, the E2E Slicer's NSMF selects the NSSMF which is responsible for the instantiation of the core network over the respective cloud/NFVI platform. The NSSMF then takes over the communication with the respective NFVO in order to launch and configure a virtual core network instance (virtual Evolved Packet Core (vEPC) network). Different core network deployment scenarios [15] are possible here, depending on the selected slice template and the related service requirements. This information dictates the exact vEPC characteristics (e.g., software images, number of VNF instances, and allocated resources), and is passed on to the NFVO, which then instantiates all the needed VNFs for the creation of the vEPC by communicating with the underlying VIM. Upon completion of the request, the NFVO notifies the E2E Slicer of the vEPC service information, such as the IP address of the MME VNF, which is used by the slicer to setup an S1AP association between the serving eNodeBs and the instantiated vEPC through the S1-Flex interface.

[1] http://www.openairinterface.org/

As soon as the S1AP association is established between the eNodeB and vEPC, the E2E Slicer employs the RAN sub-slice's NSSMF to determine an updated resource sharing policy to apply to the eNodeBs taking part in the slice, which will be taking into account the radio resource requirements of all coexisting slices per eNodeB. In our implementation, as shown in Fig. 4, after computing the new RAN resource sharing, the slicer's NSSMF will communicate with the FlexRAN controller over its NBI, and the controller will in turn enforce it by communicating with each involved eNodeB's FlexRAN agent. After this procedure is completed, the slice instance is ready for utilization.

During the slice's lifetime, which spans until a predefined time specified in the slice creation request or until its explicit termination by the slice owner over the E2E Slicer's NBI, the E2E Slicer collects runtime monitoring information to perform slice lifecycle management operations. Retrieving monitoring data and performing such management tasks per sub-slice is delegated to the respective NSSMFs. Beyond enhancing RAN and core network performance via appropriate resource scaling and reconfiguration actions, such runtime information can also help the E2E Slicer operator to better understand the overall short- and long-term system behavior and optimize its orchestration. Tools from artificial intelligence and data analytics could be used to this end, which we consider a critical topic for future research.

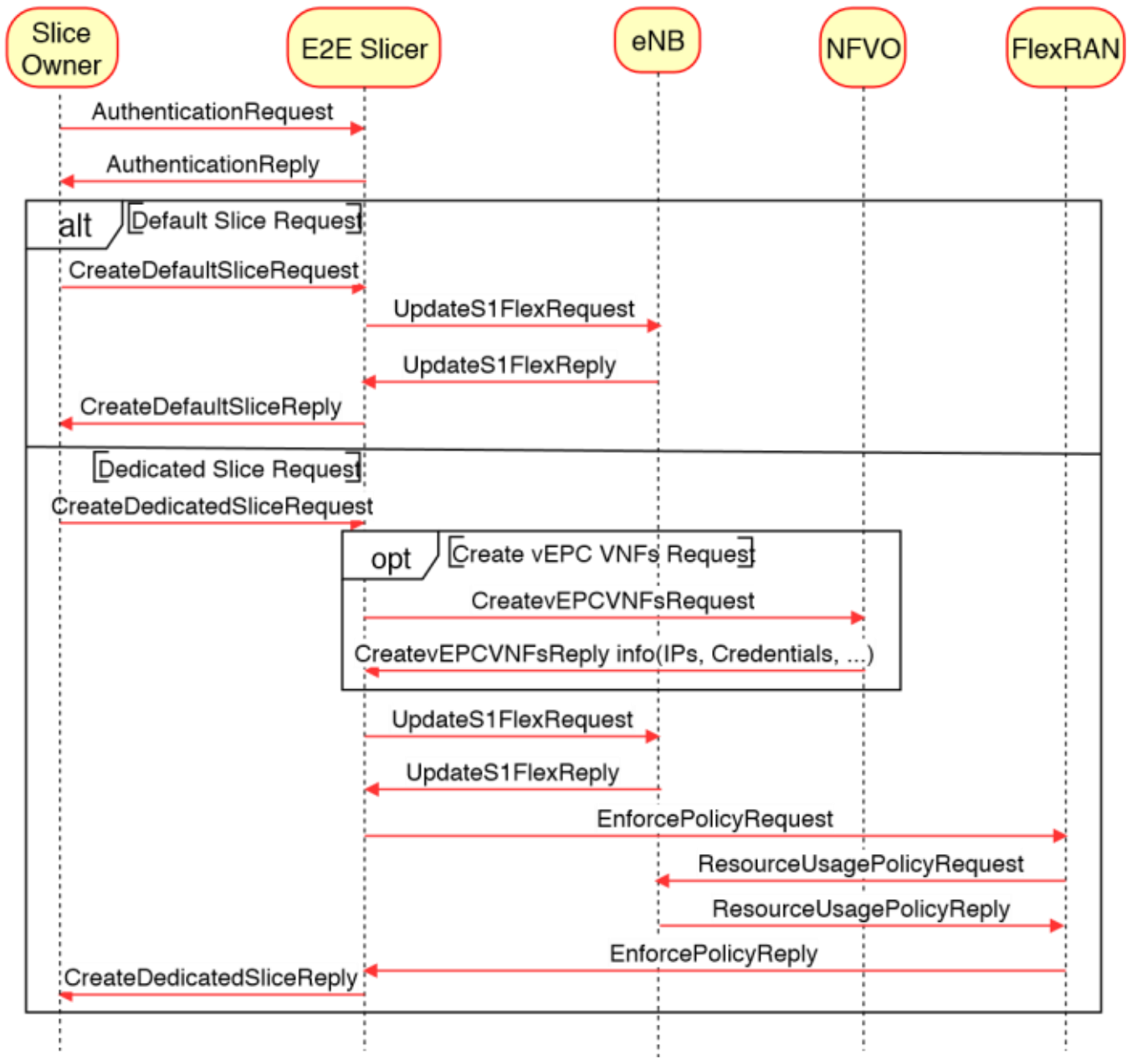

Fig. 4: Sequence diagram for the slice instantiation procedure in our implementation.

## IV. Evaluation

The results derived from the evaluation of our E2E slice orchestration platform are presented in this section. Our aim is to quantify the price to pay for the flexibility offered by our slicing design in terms of latency overhead and provide guidelines on how the system operator could use our performance results to customize the operation of the deployed slices. We have carried out our evaluation from two major perspectives: (i) a system-centric perspective which pertains to the overheads associated with slice instantiation and (ii) a user-centric one, which aims to quantify the overhead imposed for end-users during the slice operation.

### *A. System-centric evaluation: Slice instantiation overhead*

From a slice management viewpoint, we are interested in measuring the delay from the moment slice instantiation is requested over the E2E Slicer's NBI until the slice is fully operational. This system-imposed latency is usually incurred once, and that is always during the process of deploying a network slice. As depicted in Fig. 4, this overhead can be broken down to the following components: (i) the time taken for the E2E Slicer to instantiate the slice-dedicated core network VNFs on the IaaS platform, (ii) the time to communicate successfully with the eNodeB to notify it of the new vEPC instance and instruct it to create a new association with it, and (iii) the delay to configure the RAN sub-slice for the participating eNodeBs, which involves accessing the NBI of the RAN controller to update the radio resources to be shared at the RAN. The overall latency overhead is dominated by (i) and is a function of the image size for the VNFs to be deployed, some IaaS configuration options, and, importantly, the core sub-slice specific reliability and other requirements. Our testbed measurements revealed that while the E2E Slicer takes a few tens of milliseconds to communicate with both the eNodeB and the RAN controller (i.e., 0.036s and 0.027s, respectively), deploying core network VNF images in the underlying IaaS platform, which in our case is OpenStack Ocata, takes orders of magnitude more time. To demonstrate this, we have carried out the instantiation of VNF images of 10 different sizes, ranging from less than 1GB to up to about 19GB as presented in Fig 5. We note that drastic performance improvements can be achieved if VNF images are already cached at the compute nodes that will host the VNFs or if LXD containers are used with the Z File System (ZFS), since the time to transfer them from the image store is eliminated. It is thus to the advantage of both the E2E Slicer and the IaaS operator (if these two entities do not coincide) to proactively cache images at least for VNFs that are expected to be frequently used.

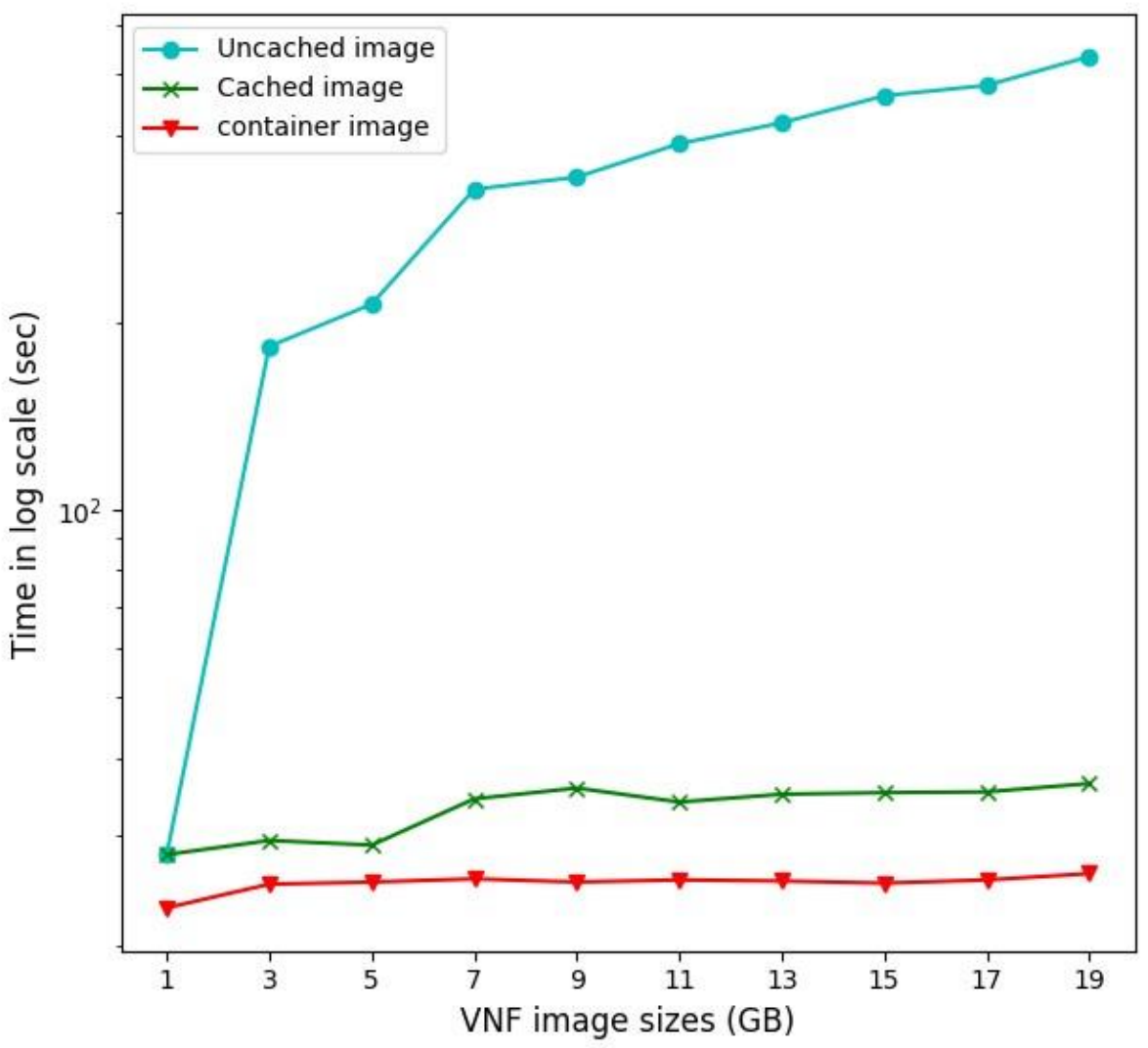

Fig. 5: VNF component instantiation times as a function of the VM and Container images size. Caching VM images at compute nodes drastically reduces the time to launch an instance.

This discussion also implies that the time to instantiate a slice and the characteristics and requirements of the latter are correlated and a trade-off for the E2E Slicer operator to address is revealed. We elaborate this by an example. Consider a vertical which requests the instantiation of a highly-available network service. In this case, and assuming a core network sub-slice which includes 4G EPC elements, the E2E Slicer will decide to launch multiple replicas of the MME VNF and appropriately balance UE association requests among them. How many virtual instances of a specific function it will launch is a matter of balancing among slice instantiation delay, available cloud resources, and the desired slice reliability levels (i.e., more VNF replicas means more time to launch them, potentially higher cost, but also increased slice robustness and availability). Measurement studies, such as the one we present here, could assist the Slicer operator with such decisions.

### *B. User-centric evaluation: Slicer scalability and UE association delays*

In our design, the E2E Slicer has a critical role in the NSSF. The level of centralization that we impose in the decision on which core network instance a UE is attached to, can increase the UE connection latency in the following two ways, which we quantify in this section:

- An additional communication exchange needs to take place anyway between the eNodeB and the E2E Slicer using the latter's REST interface upon UE attachment. It is the E2E Slicer that executes the core sub-slice selection function and drives the UE towards the appropriate core network instance.
- Compared to the default 4G core network selection decision, which is carried out in a distributed fashion by eNodeBs, in our design the E2E Slicer centrally assumes this role. It is therefore susceptible to increased latency when the request load at its end increases.

In this section, we aim to measure the effect of the above characteristics on UE attachment times. First, we report on the minimum overhead that we impose in this procedure. Compared to the default 4G attachment procedure, the additional message exchange between the eNodeB and the E2E Slicer causes an additional 36*ms* latency approximately, under nonloaded E2E Slicer conditions. This additional time is quite negligible and can be accommodated within the maximum allowable time (*T*3410 = 15*s*), which is a timer normally started by the UE at the point of attachment to the network as specified in 3GPP TS24.301. However, as the attach request workload increases, this latency is expected to also increase. We therefore carried out the following experiment. We launched the E2E Slicer as a virtual instance on an OpenStack cloud, to which we assigned a varying number of virtual CPUs from 2 to 10 (2, 4, 6, 8 and 10). Then, we generated parallel HTTP requests towards the E2E Slicer's UE attachment API endpoint and measured its response times. As Fig. 6 shows, after a specific request workload, the E2E Slicer's average response times steeply increase. For example, when it serves a workload of more than 140 requests/s, the response time goes significantly beyond 1s for all the compared number of CPUs except for 10 CPUs. This behaviour is most visible especially when the E2E Slicer runs on at most 4 CPUs.

The operator of an E2E Slicer will be certainly interested in a result such as this, since it enables the operator to adequately provision needed resources for the E2E Slicer in order to fulfill the SLA requirements of slice consumers. For example, a slice consumer might encode a specific UE attachment time constraints in the SLA requirements of the slice. The operator then has to appropriately manage the CPU and network resources allocated to the E2E Slicer to guarantee this performance objective. In our case, if the slice consumer sets a 1.5s upper bound to the average UE attachment time, the operator needs to monitor the request workload on the NSSF component and, if an increasing load is detected, the operator shall scale up/out the allocated resources to maintain a response time lower than the upper bound. For instance, this result reveals that by scaling up the number of CPUs powering the E2E Slicer from 4 to 8 when the number of connecting UEs increase beyond 140 requests/s, about 2s response delay can already be saved. This is a significant amount of delay when considering the stringent time constraint imposed by many 5G vertical applications.

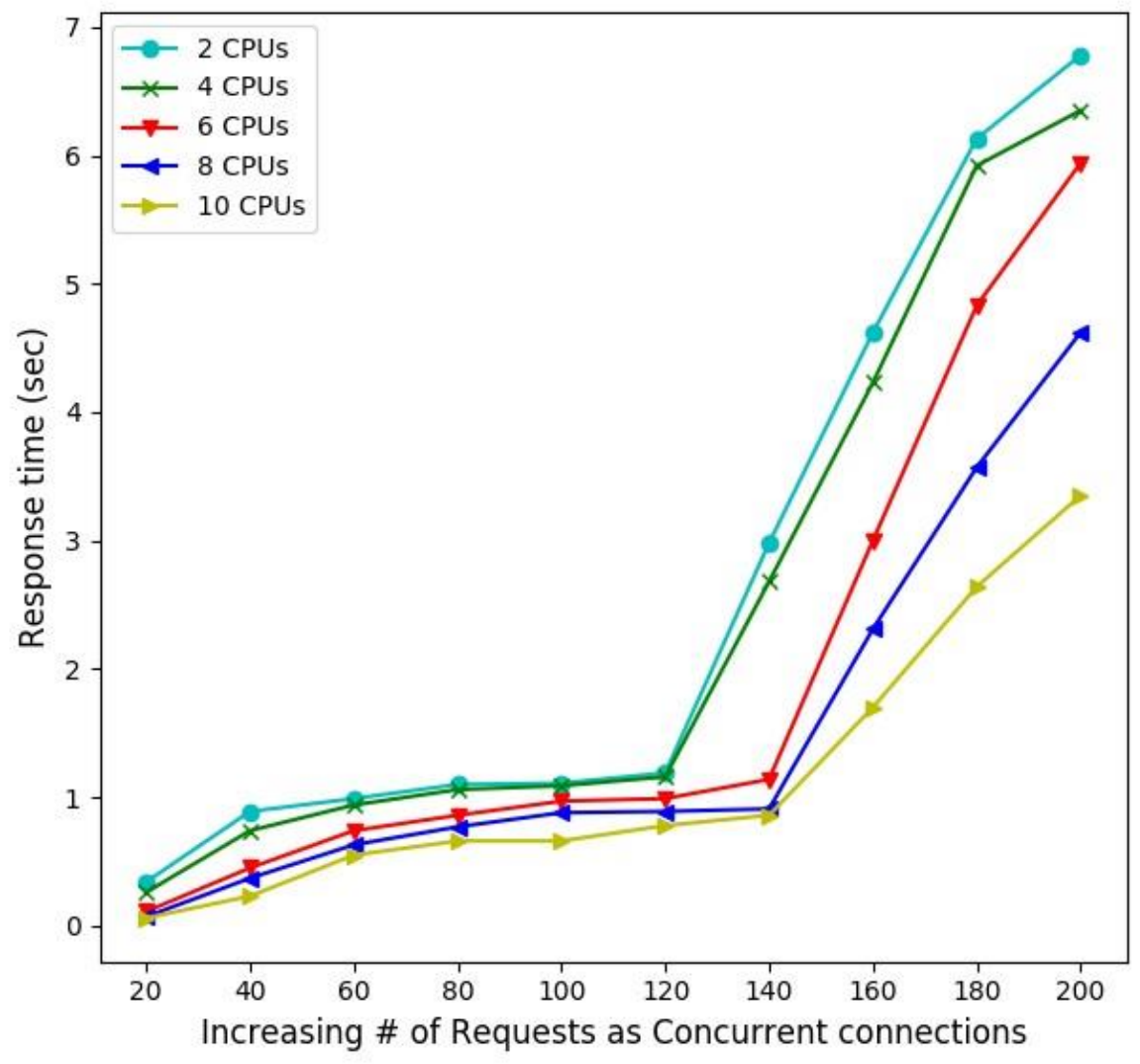

Fig. 6: Response time as a function of concurrent connections for increasing request workloads while varying the allocated CPU.

## V. Conclusion

In this article, we described the design and implementation of an architecture for E2E mobile network slicing towards supporting 5G vertical services with heterogeneous requirements in a flexible and customizable way. Experimenting with our software implementation has given us insights into how the set of developed functions can be better utilized for the overall performance enhancement of the system and how we can learn from them to better prepare the orchestration platform to take proactive measures in terms of the provisioning of system resources for network slice instances. Our results have shed light onto the limitations as well as the capacities of our E2E network slicer with respect to how fast a network slice can be instantiated from an E2E point of view and what is the performance cost from a system and user perspectives that is traded off for the achieved flexibility. Such results can be used as a basis to derive best practices for the operator of such a system.

## Acknowledgment

This work was partially funded by the Academy of Finland Project CSN - under Grant Agreement 311654 and the 6Genesis project under Grant No. 318927. The work is also partially supported by the European Union's Horizon 2020 research and innovation program under the 5G!Pagoda project and the MATILDA project with grant agreement No. 723172 and No. 761898, respectively.

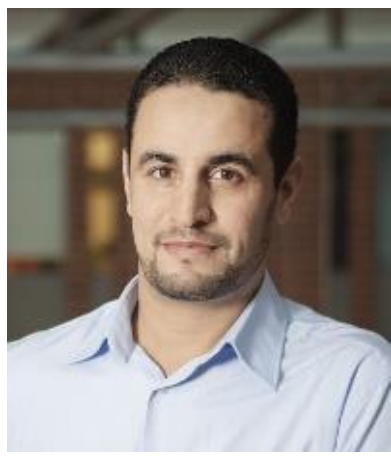

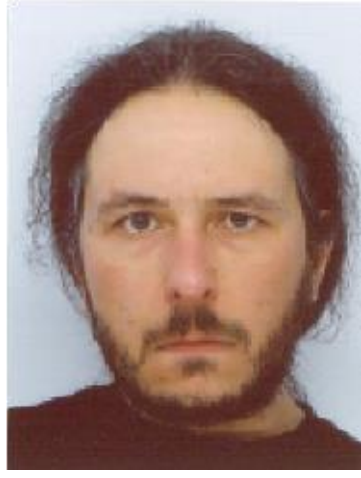

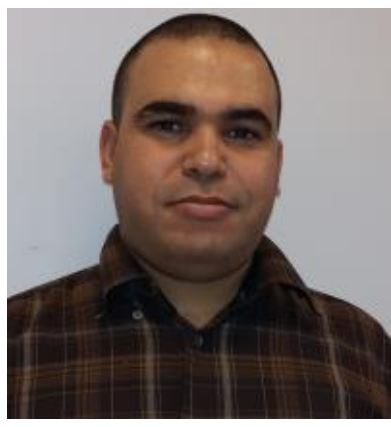

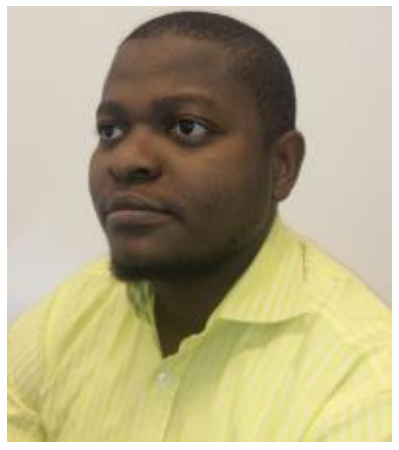

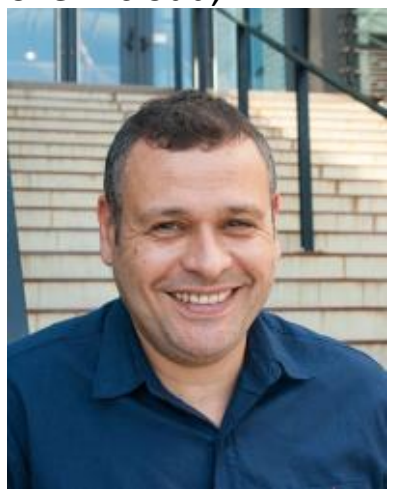

Ibrahim Afolabi obtained his Bachelors degree from VAMK University of Applied Sciences, Vaasa, Finland, in 2013 and his Masters degree from the School of Electrical Engineering, Aalto University, Finland in 2017. He is presently pursuing his doctoral degree at the same university where he obtained his Masters degree from and his research areas include Network Slicing, MEC, network softwerization, NFV, SDN, and dynamic network resource allocation.

Tarik Taleb received the B.E. degree (with distinction) in information engineering in 2001, and the M.Sc. and Ph.D. degrees in information sciences from Tohoku University, Sendai, Japan, in 2003, and 2005, respectively. He is currently a Professor with the School of Electrical Engineering, Aalto University, Espoo, Finland. He is the founder and the Director of the MOSA!C Lab. He is the Guest Editor-in-Chief for the IEEE JSAC series on network Softwarization and enablers.

Pantelis A. Frangoudis received the BSc (2003), MSc (2005), and PhD (2012) degrees in Computer Science from the Department of Informatics, AUEB, Greece. Currently, he is a post-doctoral researcher at team Dionysos, IRISA/INRIA Rennes, France, which he joined under an ERCIM post-doctoral fellowship (2012-2013). His research interests include wireless networking, Internet multimedia, network security, future Internet architectures, cloud computing, and QoE monitoring and management.

Miloud Bagaa received the bachelors, masters, and Ph.D. degrees from the University of Science and Technology Houari Boumediene Algiers, Algeria, in 2005, 2008, and 2014, respectively. He is currently a Senior Researcher with the Communications and Networking Department, Aalto University. His research interests include wireless sensor networks, the Internet of Things, 5G wireless communication, security, and networking modeling.

Adlen Ksentini received the MSc degree in telecommunication and multimedia networking from the University of Versailles, France, and the PhD degree in computer science from the University of CergyPontoise, France, in 2005. From 2006 to 2015, he was an associate professor at University of Rennes 1, France,and member of the INRIA Rennes team Dionysos. His interests include future Internet networks, mobile networks, QoS, QoE, performance evaluation, and multimedia transmission.